\documentstyle[graphicx]{mn}
\newcommand{\ls}
 {\mathrel{\hbox{\rlap{\hbox{\lower4pt\hbox{$\sim$}}}\hbox{$<$}}}}
\newcommand{\gs}
 {\mathrel{\hbox{\rlap{\hbox{\lower4pt\hbox{$\sim$}}}\hbox{$>$}}}}
\newcommand{\arcm}{\hbox{$^\prime$}}
\newcommand{\arcs}{\hbox{$^{\prime\prime}$}}
\newcommand{\et}{et al.\ }
\newcommand{\rosat}{{\it ROSAT}}
\newcommand{\euve}{{\it EUVE}}
\newcommand{\heao}{{\it HEAO-1} A2}

\title[The WFC Extragalactic Survey]
	{The ROSAT Wide Field Camera Extragalactic Survey}
\author[R.\ Edelson \et]
	{R.\ Edelson$^{1,2}$, S.\ Vaughan$^1$, R.\ Warwick$^1$,
	E.\ Puchnarewicz$^3$, I.\ George$^{4,5}$\\
$^1$X-Ray Astronomy Group; Department of Physics and Astronomy; Leicester 
	University; Leicester LE1 7RH; U.K.\\
$^2$Department of Physics and Astronomy; University of California, 
	Los Angeles; Los Angeles, CA 90095-1562; U.S.A.\\
$^3$Mullard Space Science Laboratory; University College London; Holmbury
	St Mary; Dorking, Surrey RH5 6NT; U.K.\\
$^4$Laboratory for High Energy Astrophysics; Code 662; NASA/Goddard Space
	Flight Center; Greenbelt, MD 20771; U.S.A.\\
$^5$Universities Space Research Association\\
}
\date{Submitted 30 October 1998}

\pagerange{\pageref{firstpage}--\pageref{lastpage}}
\pubyear{1999}

\begin{document}

\maketitle

\label{firstpage}

\begin{abstract}
We report the results of a new analysis of the \rosat\ Wide Field 
Camera (WFC) all-sky survey data, designed to detect extragalactic 
sources of extreme ultraviolet (EUV) radiation in regions of low 
Galactic $N_H$.
We identify a total of 19 active galactic nuclei (AGN), more than double
the number of confirmed AGN in the published WFC~(2RE) survey.
Our sample contains 8 narrow-line Seyfert 1 galaxies, making this the 
first reasonably-sized, complete sample of these extreme AGN, along with 6 
broad-line Seyfert 1s and 5 BL Lacertae objects. 
These EUV-selected sources typically have steep soft X-ray spectra with a mean 
power-law energy index  $ \alpha_{X}\approx -2$.  
The derived AGN luminosity function is rather flat and appears to cut off above 
a luminosity of $10^{46}$~erg/s/keV, measured monochromatically at 200 eV. 
Narrow-line Seyfert~1s account for roughly half of the local 
($ z \approx 0 $) volume emissivity in the EUV band.

\end{abstract}


\section{Introduction}

It is becoming increasingly clear that the extreme ultraviolet (EUV) regime 
can provide unique insight into the phenomenon of active galactic nuclei 
(AGN).
Consider, for instance, the emerging subclass of AGN known as narrow-line 
Seyfert 1 galaxies (NLS1s).
The extreme properties of NLS1s have prompted comparison with Galactic 
black hole candidates in their high states (Pounds, Done \& Osborne 1995) 
leading to the suggestion that they may represent a class of AGN in which 
accretion (onto a supermassive black hole) proceeds at a rate close to the 
Eddington limit (Ross, Fabian \& Mineshige 1992).
These objects were first identified (Osterbrock \& Pogge 1985; Goodrich 
1989) by virtue of the fact that their optical permitted lines are much 
narrower (H$\beta$ FWHM $ \ls 2000 $~km/s) than in ``normal" broad-line 
Seyfert 1 galaxies (BLS1s).
However, more recently, largely on the basis of \rosat\ observations, NLS1s 
have also been shown to exhibit extremely soft X-ray spectra and to be 
highly variable in the soft X-ray band (Boller, Brandt \& Fink 1996).
This, in turn, suggests that an excellent way to construct a representative 
sample of NLS1 objects might be via EUV-selection.

Unfortunately, the difficulties of working in at EUV wavelengths have until 
now prevented the definition of complete and unbiased samples of EUV-bright 
extragalactic sources.
These difficulties include the relative insensitivity of instruments 
flown to date compounded by the fact that most lines of sight to the 
extragalactic Universe are opaque in the EUV due to photoelectric 
absorption in the interstellar medium of our Galaxy.

During 1990, \rosat\ performed an all-sky survey in both the 
soft X-ray (0.1--2.4~keV) and EUV (60--210~eV) bands.
The former employed the \rosat\ X-ray telescope and position sensitive 
detector (PSPC; Tr\"{u}mper \et 1991) and the latter the coaligned UK Wide 
Field Camera (WFC; Sims \et 1990).
The result has been the publication of the \rosat\ Bright Source Catalogue 
(RBSC; Voges \et 1996) and the WFC RE and 2RE catalogues (Pounds \et 1993; 
Pye \et 1995). 
Here we utilize the WFC all-sky survey database to produce the first reasonably 
large and complete sample of EUV-selected extragalactic objects, all of which 
are identified as AGN. 

The remainder of this paper is organised as follows. We first discuss the 
selection criteria used to define a preliminary list of EUV-selected sources 
and then describe how this sample divides into AGN and Galactic 
stellar sub-populations. 
Next we consider the properties of the EUV-bright AGN including a comparison 
with a hard X-ray selected AGN sample. 
In \S~4 we derive an approximate luminosity function for EUV-selected AGN and 
calculate the local volume emissivity of such sources. 
Finally, in \S~5, we briefly summarise our results and consider possible future 
extensions of this work.

\section{The Source Catalogue}

The selection criteria we have used in order to include a source in our 
preliminary source list is as follows: \\
(1) A WFC all-sky survey S1 band (90--210 eV; 60--140~\AA) detection at 
$ \ge 2.5 \sigma $; \\ 
(2) An S2 band (60--110 eV; 110--210~\AA) {\it null} detection; \\ 
(3) A coincident strong RBSC X-ray detection (0.1--2.4~keV count rate  
$\ge$0.3~ct/sec) within 100\arcs\ of the S1 position; \\ 
(4) A value of the foreground Galactic column density in the direction of 
the source of $ N_H \le 2.5 \times 10^{20} $~cm$^{-2}$ (Dickey \& Lockman 
1990).

Criterion (3) allows the current survey to go more than a factor of 2 
below the 5.5$\sigma$ WFC 2RE survey limit (Pye \et 1995), since the 
requirement for {\it both} an S1 and soft X-ray detection limits the 
number chance coincidences in the sample to much less than one.
Criteria (2) and (4) select against Galactic sources, since there is  no 
reasonable expectation of extragalactic sources being detectable in the S2 
band anywhere in the sky or in the S1 band if the Galactic column is too 
high. 
For example, in the S2 band, even for the lowest-column source in the sample 
(which has $ N_H = 6 \times 10^{19} $~cm$^{-2}$), $ \tau_{S2} = 3.7 $, 
corresponding to a fractional transmission of only 2.6\%. 
For comparison the transmission in the S1 band is 16\% for this same source. 
However, at the survey limit of $ N_H = 2.5 \times 10^{20} \rm~cm^{-2}$, even 
the S1 transmission has declined to 0.6\%.
Clearly only the brightest extragalactic sources would be able to be 
detected in the presence of such strong attenuation. 

A total of 34 WFC sources were found to satisfy the above criteria.
For this preliminary sample of EUV sources, SIMBAD and other catalogues  
were searched  for potential optical counterparts.
This process proved efficient in that all but one of the sources 
(RX~J0437--47; see \S~2.2.) were identified in  this fashion.
The sample is now fully identified and comprises 19  extragalactic and 15 
Galactic sources.
Details of these sources are presented below.

\subsection{The Extragalactic Sources}

Table~1 contains a compilation of the \rosat\ data for the extragalactic 
sample as detailed in the table footnote. Note that the two hardness ratios 
$HR1$ and $HR2$ apply to the full PSPC band and the hard PSPC band 
respectively  (i.e., $ HR1 = (H-C)/(H+C)$ and $ HR2 = (H2-H1)/(H1+H2) $ 
where $C$ is the 0.1--0.4~keV count rate, $H$ the 0.5--2~keV count rate,  
$H1$ the 0.5--0.9~keV count rate, and $H2$ the 0.9--2~keV count rate).
Also we quote the RBSC positions in preference to the WFC positions
since the former are more accurate (the X-ray telescope has better 
spatial resolution and typically records at least 10 times more counts
than the WFC).

\begin{table*}
 \centering
 \caption{The WFC AGN Sample: \rosat\ Data}
 \begin{tabular}{@{}lrrrrrrrrr@{}}

     & S1  & S2 & RBSC & & & & R.A. & Dec. & Offset \\
Name & (ct/ks) & (ct/ks) & (ct/s) & HR1 & HR2  
	& $\alpha_{X}$ & (J2000) & (J2000) & (\arcs) \\ 
(1) & (2) & (3) & (4) & (5) & (6) & (7) & (8) & (9) & (10) \\ \\
WPVS 7        & 23$\pm8$ & $<$15 & 0.96$\pm0.07$ & $-0.97\pm0.01$ &  
	---            & --4.8 & 00 39 15.6 &--51 17 01 & 47\\
LB~1727      & 10$\pm4$ & $<$4  & 4.56$\pm0.21$ & $-0.52\pm0.08$ &
        $0.68\pm0.16$ & --2.5 & 04 26 01.6 &--51 12 01 & 34\\
RX J0437--47  & 19$\pm7$ & $<$22 & 0.85$\pm0.09$ & $-0.64\pm0.07$ &
	$0.33\pm0.25$ & --2.5 & 04 37 26.6 &--47 11 18 & 60\\
GB 1011+49    & 12$\pm4$ &  $<$4 & 1.94$\pm0.07$ & $-0.38\pm0.03$ &
       $-0.03\pm0.06$ & --1.2 & 10 15 04.3 & 49 26 04 & 38\\
1ES 1028+511  & 14$\pm4$ &  $<$8 & 4.46$\pm0.09$ & $-0.26\pm0.02$ & $
         0.09\pm0.03$ & --1.2 & 10 31 18.6 & 50 53 40 & 21\\
RE J1034+39   & 23$\pm5$ & $<$16 & 2.66$\pm0.09$ & $-0.74\pm0.02$ &
       $-0.37\pm0.08$ & --2.0 & 10 34 38.7 & 39 38 34 & 31\\
EXO 1055+60   & 13$\pm4$ & $<$11 & 0.39$\pm0.03$ & $-0.76\pm0.04$ &
       $-0.09\pm0.21$ & --2.0 & 10 58 30.1 & 60 16 02 & 37\\
Mkn 421       & 85$\pm9$ & $<$16 & 26.60$\pm0.23$ & $-0.21\pm0.01$ &
        $0.02\pm0.01$ & --1.2 & 11 04 27.1 & 38 12 31 & 43\\
IC 3599       & 17$\pm4$ &  $<$5 & 5.10$\pm0.11 $ & $-0.63\pm0.01$ &
       $-0.46\pm0.04$ & --1.9 & 12 37 41.4 & 26 42 29 & 96\\
IRAS 1334+243 & 15$\pm4$ &  $<$7 & 2.53$\pm0.09 $ & $-0.65\pm0.02$ &
        $-0.13\pm0.08$& --2.0 & 13 37 18.8 & 24 23 06 & 41\\
PG 1415+451   &  8$\pm2$ &  $<$8 & 0.50$\pm0.03$ & $-0.66\pm0.03$ &
       $-0.19\pm0.11$ & --2.0 & 14 17 00.5 & 44 55 56 & 34\\
NGC 5548      & 12$\pm4$ &  $<$5 & 4.95$\pm0.11$ & $-0.11\pm0.02$ &
        $0.17\pm0.03$ & --1.3 & 14 17 59.6 & 25 08 17 & 50\\
1H 1430+423   & 18$\pm5$ & $<$12 & 4.20$\pm0.09$ & $-0.60\pm0.05$ &
        $0.12\pm0.03$ & --2.0 & 14 28 32.0 & 42 40 28 & 69\\
Mkn 478       & 66$\pm7$ & $<$18 & 5.78$\pm0.10$ & $-0.70\pm0.01$ &
       $-0.11\pm0.05$ & --2.0 & 14 42 07.7 & 35 26 32 & 35\\
RX J1618+36   & 10$\pm4$ &  $<$9 & 0.85$\pm0.03$ & $-0.43\pm0.03$ &
       $-0.01\pm0.07$ & --1.6 & 16 18 09.2 & 36 19 50 & 93\\
RX J1629+40   & 12$\pm3$ &  $<$7 & 0.78$\pm0.03$ & $-0.79\pm0.02$ &
        $0.00\pm0.13$ & --2.3 & 16 29 01.2 & 40 07 53 & 29\\
PKS 2155--304 & 180$\pm12$ & $<$17 & 36.20$\pm1.53$ & $-0.42\pm0.03$ &
      $-0.04\pm0.07$ & --1.8 & 21 58 52.2 &--30 13 37 & 11\\
NGC 7213      & 12$\pm4$ &  $<$4 & 3.94$\pm0.27$ & $0.24\pm0.06$ &
       $0.23\pm0.08$ & --0.9 & 22 09 16.6 &--47 10 02 & 21\\
RE J2248--51  & 18$\pm5$ & $<$10 & 2.24$\pm0.17$ & $-0.65\pm0.05$ &
       $0.17\pm0.17$ & --2.2 & 22 48 41.4 &--51 09 51 & 26\\
  \end{tabular}
\medskip

\raggedright
The columns provide the following information: (1) the source name;
(2) the WFC S1 count rate and 1$\sigma$ error; (3) the 2.5$\sigma$ upper 
limit on the WFC S2 count rate; (4) the RBSC 0.1--2.4~keV count rate;
(5) and (6) the $HR1$ and $HR2$ hardness ratios; (7) the X-ray 
spectral index, $\alpha_X$ (see \S~3.3); (8) and (9) the right 
ascension and declination (J2000) as tabulated in the RBSC; (10) the 
difference between the RBSC and WFC positions in arcsec.

\end{table*}

Information from other wavebands is presented in Table~2: 
column~3 gives the Galactic value of $N_H$ derived from the survey of 
Dickey \& Lockman (1990), 
columns~4--6 give the V-band optical magnitude, redshift and H$\beta$  
line width, as reported in the optical observations referenced in  
column~9. Columns~7 and 8 give the monochromatic luminosity at 200~eV
($L_{200}$) and optical/EUV spectral slope, $\alpha_{OE}$, as derived 
in \S~4.\ and \S~3.3., respectively.

\begin{table*}
  \caption{The WFC AGN Sample: Derived Data and Data from Other Wavebands}
  \begin{tabular}{@{}llccccccc@{}}
 & & $N_{H}$ & V-band & & H$\beta$ FWHM & log(L$_{200}$) & & \\
Name & Type & (10$^{20}$ cm$^{-2}$) & Magnitude & Redshift &
	(km/s) & (erg/s/keV) & $\alpha_{OE}$ & Reference \\
     (1)      &  (2)   &(3)  & (4)  & (5)   & (6)  & (7) & (8) & (9)\\
              &        &     &      &       &      &     &        &\\
WPVS 7        & NLS1   & 2.5 & 14.8 & 0.029 & 1200  & 45.31 & --0.66 & 1 
\\
LB~1727       & BLS1   & 2.0 & 14.3 & 0.103 & 3460  & 45.78 & --1.10 & 2
\\
RX J0437--47  & BLS1   & 1.8 & 15.3 & 0.050 & 4600  & 45.23 & --0.82 & 3 
\\
GB 1011+49    & BL Lac & 0.8 & 16.1 & 0.200 & ---   & 45.56 & --1.16 & 4 
\\
1ES 1028+511  & BL Lac & 1.2 & 17.0 & 0.361 & ---   & 46.55 & --0.76 & 5 
\\
RE J1034+39   & NLS1   & 1.0 & 15.6 & 0.042 & 1500  & 44.59 & --1.03 & 6 
\\
EXO 1055+60   & NLS1   & 0.6 & 17.3 & 0.149 & 2030  & 45.10 & --1.01 & 1 
\\
Mkn 421       & BL Lac & 1.4 & 13.5 & 0.030 & ---   & 45.19 & --1.00 & 7 
\\
IC 3599       & NLS1   & 1.3 & 16.5 & 0.020 & 1200  & 44.06 & --0.79 & 8 
\\
IRAS 1334+243 & NLS1   & 1.2 & 15.0 & 0.107 & 2100  & 45.41 & --1.16 & 9 
\\
PG 1415+451   & BLS1   & 1.2 & 15.7 & 0.114 & 2620  & 45.20 & --1.16 & 10 
\\
NGC 5548      & BLS1   & 1.7 & 13.1 & 0.017 & 5900  & 44.06 & --1.41 & 11 
\\
1H 1430+423   & BL Lac & 1.4 & 16.4 & 0.129 & ---   & 45.82 & --0.75 & 12 
\\
Mkn 478       & NLS1   & 1.0 & 14.5 & 0.077 & 1400  & 45.59 & --1.02 & 13 
\\
RX J1618+36   & NLS1   & 1.3 & 16.8 & 0.034 &  840  & 44.29 & --0.84 & 1 
\\
RX J1629+40   & NLS1   & 0.9 & 19.0 & 0.272 & 1500  & 43.83 & --0.53 & 14 
\\
PKS 2155--304 & BL Lac & 1.7 & 13.0 & 0.116 & ---   & 46.94 & --0.83 & 15 
\\
NGC 7213      & BLS1   & 2.1 & 10.5 & 0.006 & 3200  & 43.41 & --1.81 & 1 
\\
RE J2248--51  & BLS1   & 1.4 & 15.0 & 0.102 & 2630  & 45.61 & --1.04 &
1 

\\
\end{tabular}
\medskip

\raggedright
{\em REFERENCES:} (1) Grupe (1996). (2) Guainazzi \et (1998). (3) This
paper. (4) Puchnarewicz \et (1992). (5) Polomski \et (1997). (6)
Puchnarewicz \et (1995). (7) Miller (1975). (8) Brandt, Pounds \&
Fink (1995). (9) Wills \et (1992). (10) Boroson \& Green (1992). (11)
Osterbrock (1977). (12) Sambruna \et (1997). (13) Gondhalekar \et
(1994). (14) Bade \et (1995). (15) Falamo, Pesce \& Treves (1993). 

\end{table*}

Since the EUV and soft X-ray data in Table~1 were gathered from a single 
mission at  the same time, correlations and colours measured from them are 
reliable. However, many of the data in Table~2 were taken from a variety of 
non-simultaneous measurements, and thus any ratios involving these 
quantities (e.g., $\alpha_{OE}$) may be influenced by any temporal variability 
of the source.

\subsection{ Notes on Individual Sources }

{\bf WPVS~7:}
With $ HR1 = -0.97 $, this NLS1 is the softest AGN in the sample.
There are almost no counts in bands $H1$ and $H2$, so the hardness ratio 
$HR2$ is not well-determined.
The foreground HI column is comparatively high at $ 2.5 \times 10^{20} 
$~cm$^{-2}$ and, after correcting for the Galactic transmission,
this is the brightest Seyfert galaxy in the sample (in terms of its
incident flux).
WPVS~7 appears to be a transient source, as subsequent pointed 
\rosat\ observations show that its X-ray flux dropped by a factor of 
$\sim$400 with respect to the survey measurement, the largest such variation 
seen for any Seyfert 1 galaxy (Grupe \et 1995).

{\bf LB~1727:}
The WFC~2RE catalogue reports the detection of an EUV source identified
with the Seyfert~1 galaxy LB~1727 (1H~0419--577) but at a position over
3\arcm\ offset from the optical counterpart and a corresponding RBSC 
detection. 
A close examination of the WFC data for this field reveals that 
the 2RE source is a fact a blend of two sources of roughly equal
brightness in the S1 band. 
These two sources are clearly resolved in pointed EUVE and \rosat\ HRI 
observations and have been identified respectively with the Seyfert galaxy 
LB~1727 (Guainazzi \et 1998; Turner \et 1999) and an AM Her star 
EUVE~J0425.6--5714 (Halpern \et 1998). 
When the effects of the source confusion are taken into account, the WFC
source associated with LB~1727 meets all the criteria defined in \S 2 and 
hence we include it in our sample.

{\bf RX~J0437--47:}
This was the only source in the sample without a published optical 
spectrum.
Low-resolution optical spectra were therefore acquired from the 
3.9m Anglo-Australian Telescope 
using the RGO and FORS spectrographs, with total integration times of 1000 
and 900 sec respectively, in $\sim$2\arcs\ seeing conditions.
The combined, flux-calibrated spectrum, shown in Figure~1, is that of a 
Seyfert 1 galaxy at a redshift of $ z = 0.051 $. 
The Balmer lines are clearly broad, with H$\beta$ FWHM = 4600~km/s, so we 
classify this source as a BLS1.

\begin{figure}
 \begin{center}
\rotatebox{-90}{\includegraphics[width=6.5cm]{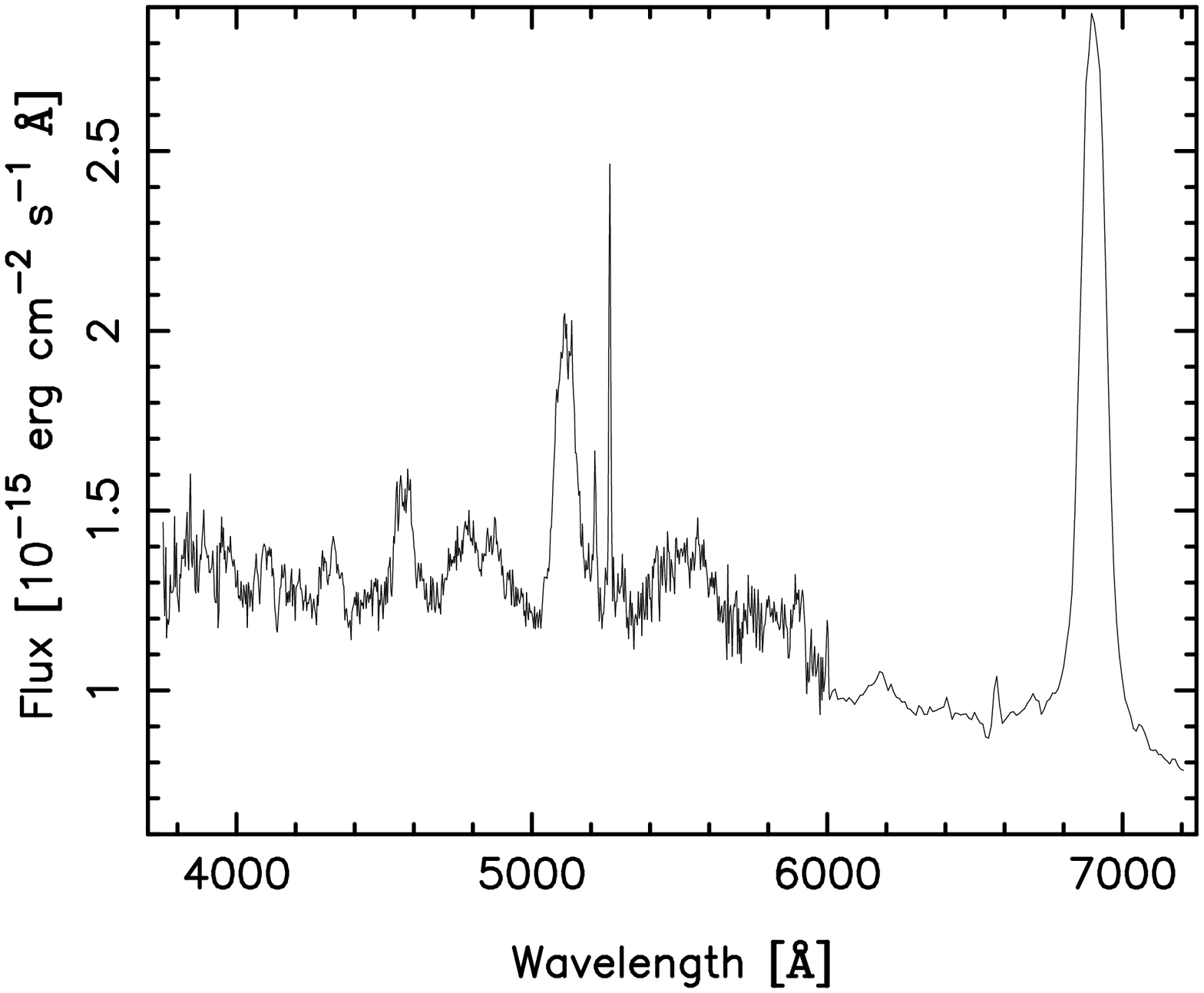}}
 \end{center}
 \caption{The optical spectrum of RX~J0437--47. }
\end{figure}

{\bf IC 3599:}
This is another highly transient source, with later pointed \rosat\ 
observations giving a flux $\sim$100 times lower than at the time of the 
all-sky survey.
A recent optical spectrum most closely resembles that of an extragalactic 
HII region, but at the time of the survey this source clearly had the
appearance of a NLS1 (Brandt, Pounds \& Fink 1995) and that classification 
is adopted herein. 

{\bf 1H 1430+423:}
This AGN is listed as an S1 and S2 detection in the WFC 2RE survey 
(Pye \et 1995) but only as an S1 detection in the original WFC RE survey 
(Pounds \et 1993). 
Our own inspection of the relevant WFC survey image confirms that any 
detection in the S2 band is at best very marginal (see Figure~2). 
The S2 detection in the 2RE processing appears to be a result of confusion 
with the high background signal in this field.
Since this source is clearly of extragalactic origin, and notwithstanding
selection criterion (2) above, we have included it in our sample.

\begin{figure}
 \begin{center}
 \includegraphics[width=4cm, height=4cm]{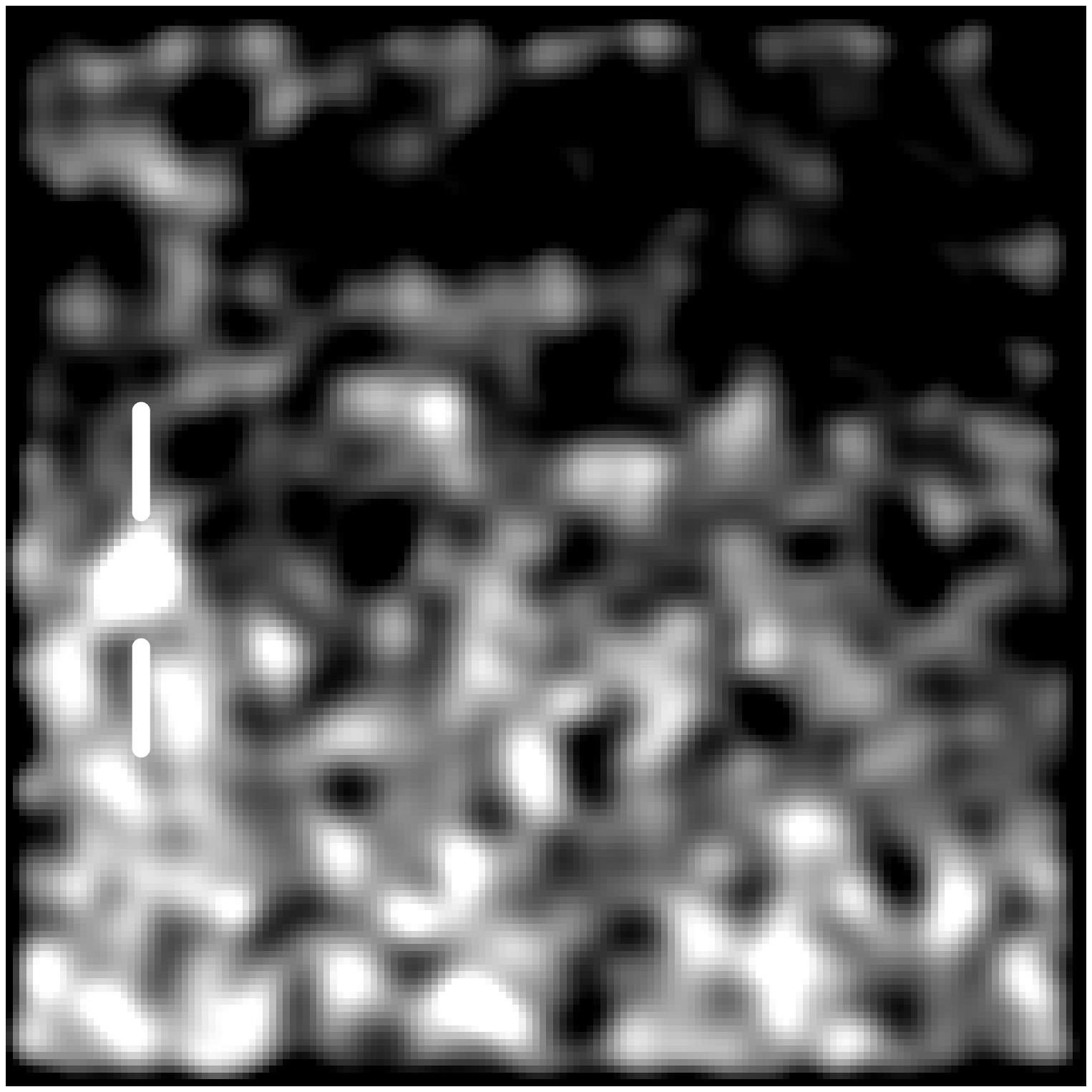}
 \includegraphics[width=4cm, height=4cm]{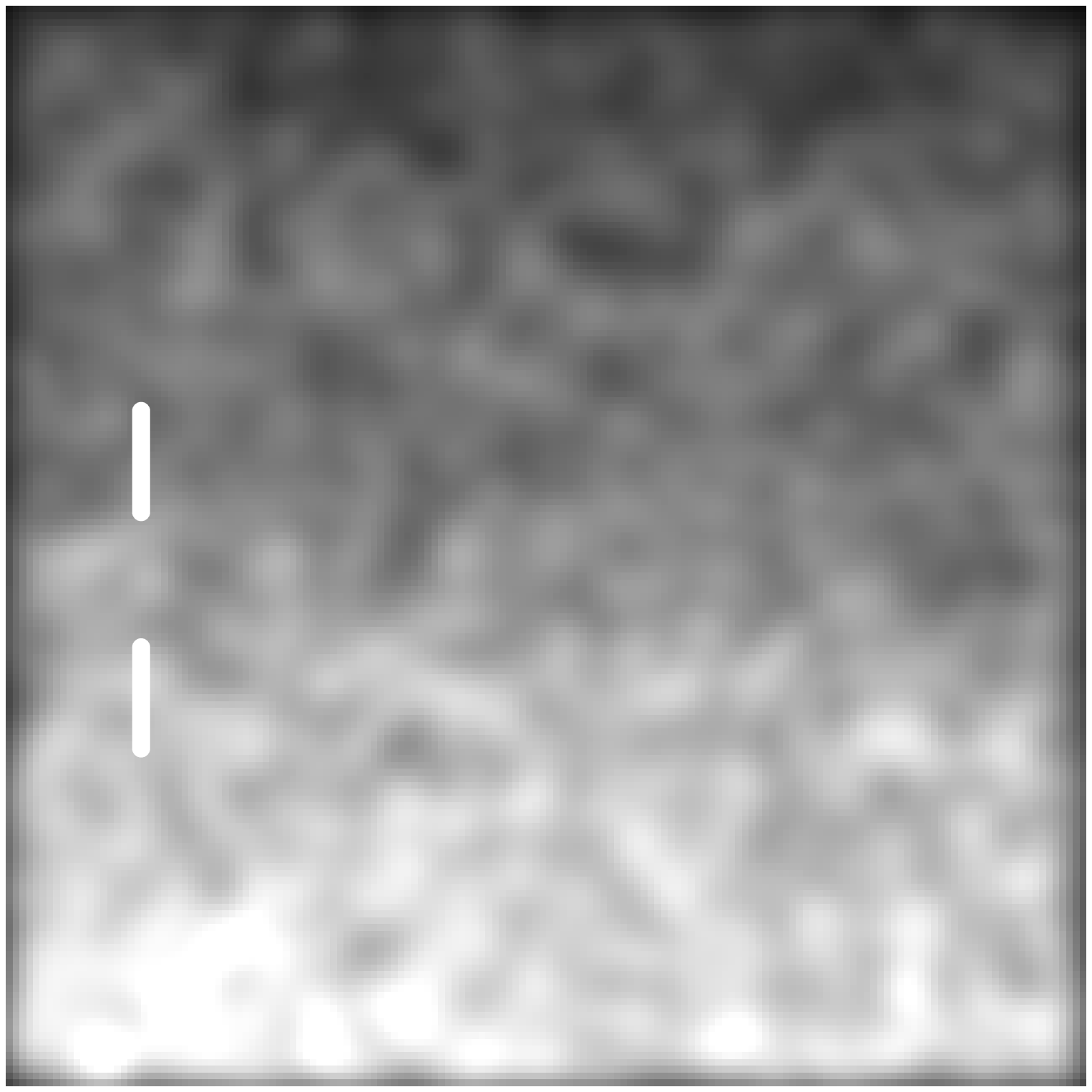}
 \caption{WFC S1 (left) and S2 (right) images of 1H 1430+423.
 The source is clearly detected (towards the left) in the S1 image but not 
 in the S2 image.}
\end{center}
\end{figure}

\subsection{The Galactic Sources }

For completeness, the Galactic sources selected by the criteria defined 
at the start of this section are listed in Table~3. 
Columns~1 and 2 give the source name and classification, 
columns 3--10 give the same quantities as in Table~1, and
column~11 gives the value of $N_H$ derived from Dickey \& Lockman (1990).
The identifications encompass 11 late-type stars, two cataclysmic 
variables (CV), a low-mass X-ray binary (XRB), and a globular cluster (GC). 
We believe that it is very unlikely that this list harbours any 
unrecognized extragalactic sources.

\begin{table*}
  \caption{WFC Galactic sample}
  \begin{tabular}{@{}llccccccccc@{}}
& & S1 & S2 & RBSC & & & R.A. & Dec & Offset & $N_{H}$ \\
Name & Type & (ct/ks) & (ct/ks) & (ct/s) & $HR1$ & $HR2$ & 
	(J2000) & (J2000) & (\arcs) &(10$^{20}$ cm$^{-2})$ \\
(1) & (2) & (3) & (4) & (5) & (6) & (7) & (8) & (9) & (10) & (11) \\ \\
V777 Tau      &  F0 & 13$\pm4$ & $<$17 & 0.56$\pm0.06$ & --0.03$\pm0.10$ &
	0.26$\pm0.13$ & 04 26 16.9 &--37 57 30 & 47 & 2.0 \\
H 0449--55     &  F5 & 41$\pm8$ & $<$30 & 1.08$\pm0.11$ & --0.20$\pm0.09$ &
	0.12$\pm0.15$ & 04 53 30.2 &--55 51 28 &  8 & 2.1 \\
HD 77407      &  G0 & 30$\pm8$ & $<$52 & 0.66$\pm0.05$ & --0.21$\pm0.07$ &
	0.01$\pm0.12$ & 09 03 26.8 & 37 50 32 & 90 & 2.4  \\
DK Leo        &  K7 & 17$\pm4$ & $<$14 & 0.51$\pm0.51$ & --0.38$\pm0.07$ &
	--0.02$\pm0.14$ & 10 14 19.3 & 21 04 37 & 43 & 2.4 \\
EK UMa        &  CV & 33$\pm6$ & $<$13 & 1.11$\pm0.04$ & --0.99$\pm0.01$ &
	--1.00$\pm0.01$ & 10 51 35.3 & 54 04 37 & 52 & 1.0 \\
CW UMa        &  M4 & 20$\pm6$ & $<$9 & 0.37$\pm0.04$ & --0.21$\pm0.10$ &
	--0.20$\pm0.16$ & 11 11 51.8 & 33 32 13 & 39 & 2.1  \\
HD 108102     &  F8 & 19$\pm4$ & $<$17 & 0.60$\pm0.04$ & --0.06$\pm0.06$ &
	--0.09$\pm0.08$ & 12 25 01.8 & 25 33 41 & 10 & 1.8  \\
SAO 632275    &  G6 & 10$\pm3$ & $<$13 & 0.39$\pm0.03$ & --0.23$\pm0.07$ &
	0.13$\pm0.12$ & 12 58 34.7 & 38 16 41 & 20 & 1.4  \\
GP Com        &  CV & 17$\pm5$ &  $<$7 & 0.54$\pm0.07$ &  0.12$\pm0.11$ &
	0.39$\pm0.13$ & 13 05 42.7 & 18 00 56 & 45 & 2.1  \\
HD 114378     &  F5 & 17$\pm5$ & $<$29 & 0.88$\pm0.06$ & --0.38$\pm0.06$ &
	--0.21$\pm0.12$ & 13 09 59.3 & 17 31 36 & 32 & 1.9 \\
NGC 5272      &  GC & 11$\pm3$ & $<$10 & 0.57$\pm0.04$ & --0.95$\pm0.02$ &
	--0.12$\pm0.56$ & 13 42 10.2 & 28 22 50 & 56 & 1.2 \\
AG +19 1315   &  K0 & 11$\pm4$ & $<$10 & 0.55$\pm0.06$ &  0.03$\pm0.09$ &
	0.41$\pm0.12$ & 14 01 58.2 & 19 25 39 & 57 & 2.4 \\
HD 131511     &  K2 & 12$\pm4$ & $<$15 & 0.36$\pm0.03$ & --0.48$\pm0.07$ &
	--0.29$\pm0.17$ & 14 53 24.3 & 19 09 15 & 65 & 2.5 \\
Her X-1       & XRB & 93$\pm8$ & $<$14 & 18.80$\pm0.14$ & --0.37$\pm0.01$ &
	0.11$\pm0.01$ & 16 57 49.6 & 35 20 33 & 33 & 1.8  \\
HD 208496     &  F3 & 16$\pm5$ & $<$37 & 0.54$\pm0.04$ & --0.04$\pm0.07$ &
	--0.31$\pm0.10$ & 21 58 31.6 &--59 00 43 & 33 & 2.1 \\
 \end{tabular}
 \medskip
\end{table*}

\section{The Extragalactic Sample of EUV-Selected Sources}

\subsection{Completeness of the Sample}

The completeness of the sample has been checked using the 
$ \langle V/V_m \rangle $ test of Schmidt (1968). 
Here $V_m$ represents the total volume of space out to the distance at which
the source flux would drop below the survey limit. 
Likewise, $V$ is the volume of space surveyed out to the actual 
distance of the object.
For a complete sample of objects, uniformly distributed in Euclidean
space, the expectation value of $ \langle V/V_m \rangle $, is 0.5 with a 
formal 1$\sigma$ error of $1/\sqrt{12N}$, where $N$ is the number of 
objects in the sample (Avni \& Bahcall 1980).

Although in practice our catalogue was selected on the basis of a
significance threshold ({\it i.e.} $2.5\sigma$ in the S1 band), 
here we make the assumption that our set of sources represents 
{\it a count rate limited sample} with a threshold (at $2.5\sigma$) of 
10 S1~ct/ksec\footnote{PG 1415+451 has a measured S1 count rate that is 
actually below our nominal survey limit. 
In this one case we simply set $V/V_m =1$.}. 
For the extragalactic sample we then obtain 
$ \langle  V/V_m \rangle  = 0.54 \pm 0.07 $, implying that the above
choice of count rate threshold is a reasonable one. The application 
of a K-S test to the distribution of the $V/V_m$ values then gives 
$P(>d) = 90\%$ (Avni \& Bahcall 1980). 
On this basis we conclude that the sample shows no clear evidence for either 
strong evolution or incompleteness.

\subsection{Comparison with Previous EUV Studies}

In total eight extragalactic sources (three NLS1 galaxies, two BLS1s
and three BL Lac objects) are listed in the \rosat\ WFC~2RE 
catalogue\footnote{A ninth extragalactic source, the normal galaxy
NGC~4787, was listed in the WFC~2RE catalogue. This identification is
almost certainly spurious as the WFC source is detected only in the S2 band.}.
The 2RE catalogue employed a combined S1+S2 significance limit of
$ \sigma_{c} \ge 5.5 $, where $ \sigma_{c} = \sqrt{\sigma_{S1}^{2} + 
\sigma_{S2}^{2}} $, and $\sigma_{S1}$ and $\sigma_{S2} $ are the 
significances of the detections in the two WFC filters.
Since AGN are very unlikely to be detected in the S2 band, 
the 2RE survey was not optimized to find such sources. 
As noted earlier, our approach here has been to search much more deeply in 
the S1 band (i.e., to $ 2.5 \sigma $), whilst guarding against spurious 
detections by requiring a simultaneous detection in the soft X-ray band 
(at a level well above the threshold of the RBSC). 
The resulting sample of extragalactic sources, all of which are identified 
as AGN, is more than twice the size of that derived from 2RE catalogue.

The Extreme Ultraviolet Explorer (\euve) has also conducted an extensive 
all-sky survey at EUV energies (Bowyer \& Malina 1991).
Using \euve\ survey data, Marshall, Fruscione \& Carone (1995) compiled a list 
of 13 extragalactic sources detected by \euve\ at  $ \ge 2.5 \sigma $, eight of 
which appear in the current WFC sample. 
However, that study cross-correlated the \euve\ data with catalogues of 
previously-known AGN. 
Since few NLS1s were known at that time, it is no surprise that this sample 
contains only three NLS1s (representing $<$25\% of the total, compared to 
almost 50\% in our current sample). 
One can therefore conclude that the \euve\ sample is probably incomplete 
and biased against NLS1s. 
A low-significance survey by Fruscione (1996) and Craig \& Fruscione (1997) 
resulted in a large number of potential EUV detections of extragalactic 
sources. 
However, as recognised by the authors, a rather high fraction of these EUV 
sources may be spurious and even the bona fide EUV detections
may represent chance coincidences with AGN (in the relatively large 
\euve\ error circles) or arise due to the hard leak in the \euve\ filters.
The smaller error circles, the S2 discriminant and the sharper filter cutoff
are all advantages of the WFC survey compared to that carried out by \euve, 
at least in the narrow context of defining a complete EUV-selected sample 
of AGN. 
It must be emphasized that {\it all} of the identifications in the 
current WFC survey are very likely to be solid and secure.

\subsection{Comparison with a Hard X-ray Selected Sample}

It is interesting to compare the properties of our sample of sources selected 
in EUV band with those of a sample of AGN selected at much harder X-ray 
energies.
For this purpose, we use the well-studied set of AGN derived from the 2--10~keV 
\heao\ survey (Piccinotti \et 1982).
Table~4 compares certain properties of the WFC and \heao\ samples.
Clearly there is quite a striking difference in the make-up of the AGN
population as one moves up roughly a factor of 25 in energy.
The \heao\ sample is dominated by BLS1s, and contains four narrow emission-line 
and Seyfert~2 galaxies, but {\it no} NLS1s.
By comparison, almost half of the AGN in the EUV-selected WFC sample
are NLS1s,
which also contains a significantly higher proportion of BL Lacs, but
{\it no} Seyfert~2s. 
This is yet another example of the relationship between optical emission 
line width and EUV/X-ray spectral properties.
The WFC objects are optically fainter and typically
are at higher redshift than the \heao\ sources. 
Since the EUV sample is selected from a smaller region of sky (effectively 
$\sim$31 square degrees; see \S~4), it is not unexpected that it is necessary to 
search to greater distances (and fainter magnitudes) in order to find comparable 
numbers of sources in the EUV band as contained in the \heao\ sample.

\begin{table}
 \centering
 \caption{Comparison of the WFC and \heao\ samples}
 \begin{tabular}{@{}lcc@{}}
                          & WFC          & \heao      \\
   \\
   NLS1                   & 8 (42\%)     & 0 (0\%)    \\
   BLS1/quasar            & 6 (32\%)     & 27(78\%)   \\
   Seyfert 2/NELG         & 0 (0\%)      & 4 (11\%)   \\
   BL Lac                 & 5 (26\%)     & 4 (11\%)   \\
   Total                  & 19 (100\%)   & 35 (100\%) \\
   \\
   $\langle z \rangle $   & 0.079        & 0.035      \\ 
   Median redshift        & 0.050        & 0.019      \\
   $\langle HR1 \rangle $ & --0.52       & +0.47      \\
   $\langle HR2 \rangle $ & --0.02       & +0.28      \\
   $\langle V \rangle $   & 15.2         & 13.5      \\
 \end{tabular}
\end{table}

As noted earlier, the RBSC provides two measures of spectral hardness 
covering the 0.1--2.0~keV band, namely the ratios $HR1$ and $HR2$. 
Figure~3 shows a plot of $HR1$ versus $HR2$, for both our EUV-selected 
sample and the \heao\ sources. 
There is evidence of a correlation, indicating that sources that are 
characterized by a significant flux in the 0.1--0.4 keV band (i.e., 
$ HR1 \ls 0 $) also have rather steep spectra in the adjacent
0.5--2.0 keV band ($ HR2 \ls 0 $), and vice-versa.
There is very little overlap in $HR1$ between the two samples, even
though neither explicitly used any colour selection criteria.
Again, this emphasizes how selection in the hard X-ray and EUV bands finds 
very different types of objects.

\begin{figure}
 \begin{center}
\rotatebox{-90}
 {\includegraphics[width=6.5cm]{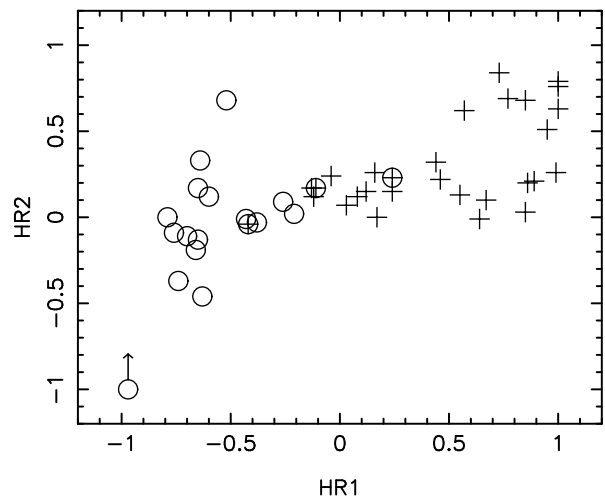}}
 \end{center}
 \caption{
 Plot of soft- and hard-band hardness ratios ($HR1$ and
 $HR2$, respectively).
 A positive hardness ratio implies a flat (hard) spectrum.
 Circles refer to objects in the WFC sample and crosses, to sources in
 the \heao\ sample. 
 Three objects are common to both samples; they have coincident   
 crosses/circles.
 The NLS1 WPVS~7 is denoted by a circle with an arrow to indicate a
 lower limit to its unmeasured {\it HR2}.
 }
\end{figure}

\subsection{The Spectral Form of the Sources}

For each WFC source, we have measurements available in four spectral 
channels, namely the WFC S1 band count rate and soft X-ray count rates 
measured by the \rosat\ PSPC in the $C$, $H1$ and $H2$ bands (derived in an 
approximate way from the broad band, 0.1--2.4 keV count rate tabulated in 
the RBSC and the corresponding $HR1$ and $HR2$ hardness ratios; see 
\S~2.1).
We have employed the spectral fitting package {\sc xspec v10.0} to model these 
four 
channel data in terms of a simple power-law continuum with absorption
corresponding to the foreground Galactic $N_H$. The energy spectral 
index, $\alpha_X$ (defined as $ S_\nu \propto \nu^\alpha$), derived for 
each source is tabulated in Table~1.

In practice, the simple spectral model detailed above gave an unacceptable 
minimum $\chi^{2}$ for six of the sources. However, when the constraints on 
the value $N_H$ were relaxed, in each case a much better 
(and acceptable) fit was obtained. 
Interestingly all of these sources required an $N_H$ in excess of the Galactic 
value in order to improve the fit (typically by an amount 
$\Delta N_H \approx 5 \times 10^{19}\rm~cm^{-2}$).
Three of these sources are BL Lac objects and it is possible to invoke 
spectral curvature (downwards below $\sim 0.3$ keV) as the cause of the 
discrepancy. 
The other sources are IC~3599 and RE~J1034+39, both of which are NLS1s, 
and LB~1727, a BLS1 galaxy. 
More detailed spectral fitting of IC~3599 (Brandt \et 1995),
RE~J1034+39 (Pounds \et 1995) and LB~1727 (Turner \et 1999) is
consistent with additional $N_H$ or a relatively complex spectral form
in these objects.
Of course, the assumed value for the Galactic $N_H$ (based on the broad-beam 
21 cm measurements compiled by Dickey \& Lockman 1990) may be in error by 
at least $N_H \sim 10^{19}\rm~cm^{-2}$ and possibly more in some instances 
(Elvis, Wilkes \& Lockman 1989). 
Nevertheless, one conclusion is clear from the above analysis, namely that we 
have no evidence for any excess flux in the extreme ultraviolet band in the 
spectra of NLS1s (over and above that predicted by a simple 
power-law extrapolation of the soft X-ray spectrum).

Using the unabsorbed fluxes measured at 200 eV and an estimate of the optical 
flux at 5500~\AA\ derived from the V-band magnitude, we have also calculated the 
optical/EUV spectral index, $\alpha_{OE}$, for each source (see Table~2).
Figure~4 shows a plot of $\alpha_{X}$ versus $\alpha_{OE}$ for the AGN 
sample. 
An obvious point is that, apart from two objects, spectral slope in the soft 
X-ray band is generally steeper than the optical/EUV index. 
Specifically $ \langle \alpha_X \rangle = -2.0 \pm 0.9 $ and 
$ \langle \alpha_{OE} \rangle = -1.0 \pm 0.3 $ (where the error represents the 
standard deviation of the distribution).
The implication is that many of these sources may contain a large hidden 
ultraviolet/EUV excess, possibly attributable to emission from the hotter 
inner regions of an accretion disk (e.g., Bechtold \et 1987).

\begin{figure}
 \begin{center}
 \rotatebox{-90}{\includegraphics[width=6.5cm]{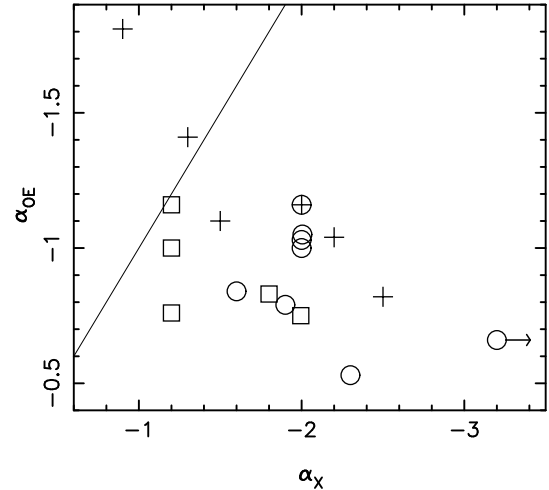}}
 \end{center}
 \caption{  Plot of the soft X-ray spectral slope, $\alpha_{X}$, versus
  the optical/EUV spectral index, $\alpha_{OE}$, for the EUV-selected AGN 
  sample.  The circles correspond to NLS1s, the crosses to the BLS1s and the 
  squares to the BL Lac objects. 
  The solid line corresponds to $ \alpha_{OE} = \alpha_X $, as would be the case
  if the source was a pure power-law.
  The extreme NLS1, WPVS 7, has
  $\alpha_{X} = -4.8$ and is denoted as the circle with an arrow 
  (for an upper limit) in the lower right corner.}
\end{figure}

\section{The Luminosity Function of EUV-Selected AGN} 

Our catalogue of EUV-selected AGN represents the best
sample of such sources presently available and, in principle, may be used 
to investigate the statistical properties of the extragalactic EUV-source 
population. 
Here we construct an approximate EUV luminosity function 
and estimate the relative contributions of the various types of 
AGN to the local ($ z \approx 0 $) EUV volume emissivity.  
Values of $ H_0 = 50 $~km/s/Mpc and $ q_0 = 0.5 $ are assumed throughout 
this section.

The first step is to define the intrinsic luminosity of each source. 
For this purpose we employ a ``standard'' spectral model consisting of a 
power-law continuum with $\alpha_X$ fixed at --2 (i.e., close to the overall 
mean value for the sample). 
We also investigated using the individual $\alpha_X$ values tabulated in Table~1 
and found that it made no significant difference, and thus stayed with the mean 
value because the uncertainties are smaller.
Using this spectral form, together with the appropriate Galactic $N_H$ and 
redshift, we then transform the observed S1 count rates for each source to a 
{\it monochromatic} luminosity, $L_{200}$, measured at a rest-frame energy of 
200 eV. 
The derived values of $L_{200}$ are listed in Table~2. 
Allowing for the strong cut-off of the source spectra at low energies due to 
Galactic absorption, the selected energy of 200 eV is close to the typical 
photon energy recorded for this set of sources in the WFC S1 band.
Similarly, we use a monochromatic luminosity since the typical effective 
detection bandwidth is very narrow (again as result of Galactic absorption). 
It is a matter of semantics as to whether the derived luminosities are referred 
to as EUV or ``ultra-soft" X-ray measurements; here we use the former.

The standard formula for calculating the luminosity function 
$\Phi(L)$ of a sample of sources (e.g., Schmidt 1968) is:
\begin{center}
 \[ \Phi(L) = \frac{1}{\Delta L} \sum_{i=1}^{N} \frac{1}{V_{max}}_{i}, \]
\end{center}
where there are $N$ objects in each luminosity bin of width $ \Delta L $,
and ${V_{max}}_i$ is the volume surveyed in the process of detecting the 
$i$th source. 
Because the $V_{max}$ test was developed for a flux limited survey, the 
calculation of $V_{max}$ is complicated in this instance by the fact that the 
WFC survey is significance limited (that is, the limiting flux is not constant 
from one part of the sky to another).
Earlier (\S~3.1), we assumed that the sample of sources could, at least to 
a first approximation, be represented as a {\it count-rate} limited sample 
with a threshold (at $2.5\sigma$) of $C_{lim} = 10$~S1~ct/ks.  
Since this prescription gave a reasonable outcome in terms of the 
$ \langle V/V_m \rangle $ test, we also adopt the same approach in 
calculating the luminosity function; in any event the very limited source 
statistics most likely dominate the errors. 

A second serious complication is that the transmission in the EUV band is a 
very strong function of the Galactic foreground $N_H$. 
Our approach of assuming a constant count rate threshold over the sky implies
a variable flux threshold (when due allowance is made for the variable Galactic 
transmission), which in turn directly influences the survey volume. 
Specifically, for each source we calculate
\begin{center}
\[ V_{max_{i}} = \int^{N_{Hlim}}_{0} \frac{1}{3}\Omega(N_{H}) (\frac{T(N_{H})}
{T(N_{H_{i}})})^{\frac{3}{2}} d_{i}^{\frac{3}{2}} 
(\frac{C_{i}}{C_{lim}})^{\frac{3}{2}}
 dN_{H}, \]
\end{center}
where $\Omega(N_{H})$ is the differential  solid angle of sky with a 
Galactic column density $N_H$, $T(N_{H})$ is the Galactic
EUV transmission as a function of $N_{H}$, $N_{H_{i}}$ is the Galactic column 
density in the direction of the source,
$N_{Hlim} = 2.5 \times 10^{20}\rm~cm^{-2}$, $C_{i}$ is the S1 count 
rate and $d_{i}$ is the source distance. 
The cumulative function of $\Omega(N_{H}) T(N_{H})^{\frac{3}{2}}$ is shown 
in Figure~5 and has a value at $N_{Hlim}$ of $ 9.5 \times 10^{-3} $ 
steradians or 31 square degrees (whereas integration of $\Omega(N_{H}$)
between the same limits gives $\sim$3.5 steradians).

\begin{figure}
 \begin{center}
  \rotatebox{-90}
 {\includegraphics[width=6.5cm]{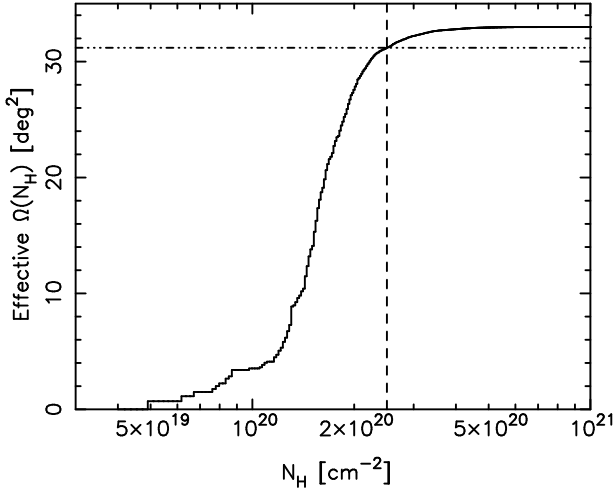}}
 \end{center}
 \caption{Cumulative form of the effective sky area function 
 $\Omega(N_{H})  T(N_{H})^{\frac{3}{2}} $. }
\end{figure}

For the purpose of this paper we have estimated the luminosity function, 
$\Phi(L_{200})$, for the full sample of AGN (i.e., Seyfert 
galaxies plus BL Lac objects) from a binned representation of the data
as shown in Figure~6. The plotted errors correspond simply to a factor 
$ 1/\sqrt{N} $ and since they don't account for any possible systematic errors, 
they should be considered to be lower limits on the true errors. 
The three lower luminosity points in Figure~6 are consistent with the 
power-law form 
\begin{center}
\[ \Phi(L_{200,44}) = 10^{-6} L_{200,44}^{-2}\rm~Mpc^{-3}~(10^{44} 
erg~s^{-1}~keV^{-1})^{-1} \]
\end{center}
where $L_{200,44}$ is the monochromatic luminosity at a rest frame 
energy of 200 eV in units of $10^{44}\rm~erg~s^{-1}~keV^{-1}$.
This rather flat luminosity function appears to cut off sharply in the 
highest luminosity bin (which contains only two sources, both of which
are BL Lac objects).

\begin{figure}
 \begin{center}
 \rotatebox{-90}
 {\includegraphics[width=6.5cm]{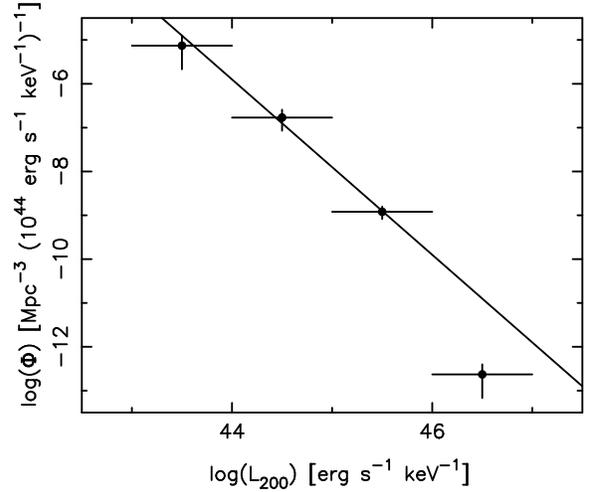}}
 \end{center}
 \caption{The EUV luminosity function of the AGN in the WFC sample.
 The monochromatic source luminosities are measured at 200 eV in the
 rest frame. The binning is logarithmic with $ \Delta$log$(L_{200}) = 1.0$.
 The solid line represents the power-law function defined in the text.}
\end{figure}

The integrated volume emissivity has been estimated by summing the
product of $1/V_{max}$ and $L_{200}$ over the sample of sources. 
We obtain a total emissivity of $6\times10^{38}\rm~erg~s^{-1}~keV^{-1}~
Mpc^{-3}$ at 200~eV (which is close to the value obtained by integrating the 
above luminosity function between $10^{43} - 10^{46} \rm 
~erg~s^{-1}~keV^{-1}$).
The NLS1s contribute 50\% of this emissivity, BLS1s 42\%, and the BL
Lac objects the remainder. 
This high contribution of  NLS1s is in contrast to the situation at hard 
X-ray energies, where NLS1s make a negligible contribution to the volume
emissivity (e.g., Picccinotti et al. 1982). 
However, it is interesting to note that a similar value 
for the volume emissivity at 200 eV is obtained by extrapolating 
downward with a spectral slope $\alpha_{X} =-0.7$ from the 2--10 keV
AGN volume emissivity derived from  the \heao\ sample (Piccinotti et al. 
1982).

\section{Conclusions }  

This paper focuses on a sample of extragalactic sources selected on
the basis of their detection in the EUV by the ROSAT WFC. 
Using an approach optimized to find such sources we construct an initial
catalogue of 34 sources, which after the exclusion of sources identified
with Galactic objects, reduces to a sample of 19 EUV-bright AGN.
This is the first reasonably sized, complete and unbiased sample of 
EUV-selected AGN to be available. 
NLS1s are well represented in the sample making up just under half of the total, 
with BLS1s and BL Lac objects comprising the remainder. 
This is in stark contrast with hard X-ray selected samples, in which NLS1s have 
negligible representation.
It is expected that these data will allow the first reliable statistical 
studies of this important emerging subclass of ``extreme" AGN.

In the first such preliminary study, presented in this paper, a  
correlation was seen between the \rosat\ soft and hard band hardness 
ratios, $HR1$ and $HR2$, indicating that sources with strong EUV excesses 
also have steeper soft X-ray spectra.
These data were also used to directly derive the first luminosity 
function for AGN measured at EUV energies (specifically at 200~eV). 
The luminosity function implies a roughly equal contribution to the EUV volume 
emissivity from each decade of luminosity for $L_{200}$ between $10^{43}$ to 
$10^{46}\rm~erg~s^{-1}~keV^{-1}$, but with a sharp cut off at higher 
luminosities. This luminosity function provides an independent estimate of 
the mean intergalactic ionizing photon field, which until now has been 
estimated by 
extrapolation from the optical luminosity function of AGN. 
Finally, we note that NLS1s contribute roughly half of the volume emissivity at 
200 eV, again in contrast to the situation pertaining at harder X-ray energies.

\section*{ Acknowledgments }

The authors would like to thank John Pye, Steve Sembay, Keith Sohl and the 
Leicester WFC team for help compiling the deep WFC data and correlating it 
with the RBSC and other data.
This research made use of data obtained 
from the High Energy Astrophysics Science Archive Research Center 
(HEASARC), provided by NASA's Goddard Space Flight Center, 
from the NASA/IPAC Extragalactic Database (NED), provided by NASA/JPL under 
contract with Caltech,
from the Leicester Database and Archive Service (LEDAS) at the Department 
of Physics and Astronomy, Leicester University, UK, and 
from the Set of Identifications, Measurements and Bibliography for 
Astronomical Data (SIMBAD), maintained by the Centre de Donnees 
astronomiques de Strasbourg. 
We thank the staff at the Anglo-Australian Observatory for obtaining the
spectrum of RX J0437--47 during service time, and Thomas Boller for a quick and 
helpful referee's report.
SV acknowledges support from PPARC.

\bsp

\label{lastpage}

\end{document}